\documentclass[aps, twocolumn, superscriptaddress]{revtex4}
\usepackage{graphicx}
\usepackage{amssymb}
\usepackage{epstopdf}
\usepackage{subfigure}

\graphicspath{{figures//}}

\begin{document}
\title{Generation of broad XUV continuous high harmonic spectra and
isolated attosecond pulses with intense mid-infrared lasers}
\author{C. Trallero-Herrero}
\affiliation{J. R. Macdonald Lab., Physics Department, Kansas State University, Manhattan, KS 66506, USA}
\affiliation{Joint Attosecond Science Laboratory, National Research Council of
Canada and University of Ottawa, Ottawa, ON K1A 0R6, Canada}
\author{C. Jin}
\affiliation{J. R. Macdonald Lab., Physics Department, Kansas State University, Manhattan, KS 66506, USA}
\author{B. E. Schmidt}
\affiliation{Institut National de la Recherche Scientifique, Varennes, Qu\'ebec J3X 1S2, Canada}
\author{A. D. Shiner}
\affiliation{Joint Attosecond Science Laboratory, National Research Council of
Canada and University of Ottawa, Ottawa, ON K1A 0R6, Canada}
\author{J-C. Kieffer}
\affiliation{Institut National de la Recherche Scientifique, Varennes, Qu\'ebec J3X 1S2, Canada}
\author{P. B. Corkum}
\affiliation{Joint Attosecond Science Laboratory, National Research Council of
Canada and University of Ottawa, Ottawa, ON K1A 0R6, Canada}
\author{D. M. Villeneuve}
\affiliation{Joint Attosecond Science Laboratory, National Research Council of
Canada and University of Ottawa, Ottawa, ON K1A 0R6, Canada}
\author{C. D. Lin}
\affiliation{J. R. Macdonald Lab., Physics Department, Kansas State University,
Manhattan, KS 66506, USA}
\author{F. L\'egar\'e}
\affiliation{Institut National de la Recherche Scientique, Varennes, Qu\'ebec
J3X 1S2, Canada}
\author{A. T. Le}
\affiliation{J. R. Macdonald Lab., Physics Department, Kansas State University,
Manhattan, KS 66506, USA}

\date{}                                           

\begin{abstract}
We present experimental results showing the appearance of a near-continuum in the high-order
harmonic generation (HHG) spectra of atomic and molecular species as the
driving laser intensity of an infrared pulse increases. Detailed macroscopic simulations reveal that
these near-continuum spectra are capable of producing IAPs in the far field if
a proper spatial filter is applied. Further, our simulations show that the near-continuum
spectra and the IAPs are a product of strong temporal and spatial reshaping
(blue shift and defocusing) of the driving field. This offers a possibility of
producing IAPs with a broad range of photon energy, including plateau
harmonics, by mid-IR laser pulses even without carrier-envelope phase
stabilization.

\end{abstract}

\maketitle


It has been a decade since the first reported generation of isolated attosecond
pulses (IAPs) \cite{Krausz:Attosecond_Metrology} and the first full
characterization of an attosecond pulse train (APT)
\cite{Agostini:Attosecond_train}, both seminal results that gave birth to the
field of attosecond science \cite{corkum:attosecond_review,
Krausz_Ivanov:Attosecond_review,Corkum_Chang:Attosecond,
KM:Attosecond_Pulses_Theory_1997,JILA:review2010}. While APTs can be now
routinely generated based on high-order harmonic generation (HHG) in gases,
production of IAPs is still a very active field of research. Common to all
methods of generating  IAPs is the concept of temporal gating, which is used to
select a XUV burst from an APT. Current methods can be divided into three
different categories. The most intuitive, but technically demanding, approach is
based on XUV spectral filtering near the cutoff of the
HHG spectra. This approach requires a short driving pulse (about two optical
cycle) with a stabilized carrier-envelope phase.
\cite{KM:Attosecond_Pulses_Theory_1997, Goulielmakis:Single_Attosecond}. This
approach could be called intensity gating, as it selects the harmonics emitted
near the peak of the most intense half-cycle of the (short) laser pulse. The second approach is based on
polarization gating \cite{Corkum:polarization_gating,
Constant:attosecond_polarization_gating, nisoli:polarization_gating}, and the
generalized double optical gating \cite{Chang:double_optical_gating,
Chang:GDOG, Zhang:GDOG_no_CEP}, in which the short driving pulse requirement is
somewhat relaxed. While these two approaches have their origin already in the
single-atom response, the third approach is based on the macroscopic
propagation effects \cite{gaarde:ATTO_filter}, which includes, ionization
gating \cite{Pfeifer_Leone:Ionization_gating, Leone:Ionization_gating_HHG,
Thomann:2009}, phase matching \cite{Marangos:HCO_2007}, and ionization-driven
reshaping Gaarde {\it et al} \cite{Gaarde:Atto_spatio_temporal,
gaarde:ATTO_filter}. It has also been demonstrated recently that near-continuum
spectra can be generated with two-color multicycle driving laser pulses by
carefully adjusting the wavelength of the supplementary pulse
\cite{Midorikawa:2color_2010}.

In order to produce a short IAP, XUV radiation with a broad frequency bandwidth needs to
be generated. Since the HHG cutoff law is $\Omega_{cutoff} \sim
I_L\lambda_L^2$, there are only two ways to extend $\Omega_{cutoff}$:
increasing the laser intensity $I_L$, or increasing the laser wavelength
$\lambda_L$. The first option is limited by ionization, which dramatically
reduces HHG yield due to the depletion of neutral atoms and, more importantly,
the phase mismatching due to free electrons in the medium. Using a short laser
pulse does not help much, since the current technology nearly reaches the
one-cycle limit already. The second option is to use lasers with a long
wavelength. This comes with an additional advantage of generating a shorter IAP
since the atto-chirp is inversely proportional to the wavelength of the driving
laser \cite{Tate:Wavelength_scaling}.  The price to pay is that HHG yields have been shown to drastically decrease with wavelength
\cite{Shiner:wavelength_scaling, Tate:Wavelength_scaling,
Starace:wavelength_scaling, Burgdorfer:Wavelength_scaling}. However, it has been demonstrated recently that this decreasing yield with laser wavelength at the single atom level can be compensated for by increasing the density of atoms resulting in increased XUV flux for some phase matching conditions
\cite{KM:Phase_Matching}.

In this paper we present experimental evidence of generation of a broad
near-continuum in HHG spectra from atomic and molecular gases by using
two-cycle mid-IR (1825 nm) laser pulses from atomic and molecular targets. As
supercontinuum emission is consistent with an IAP, we carry out detailed
theoretical simulations to show that this scheme indeed is capable of producing
IAPs if a proper spatial filter is applied. Our analysis shows that the
generation of a near-continuum and an IAP is the result of combined effect of
reshaping of the driving laser and non-adiabatic phase mismatch, caused by the
free electrons due to sub-cycle strong field ionization. With this new method and the light source used,
it is possible to use a broad range of harmonics, including plateau harmonics,
to generate an IAP with a tunable XUV photon energy.

\begin{figure}
\centering
\includegraphics[scale=0.18]{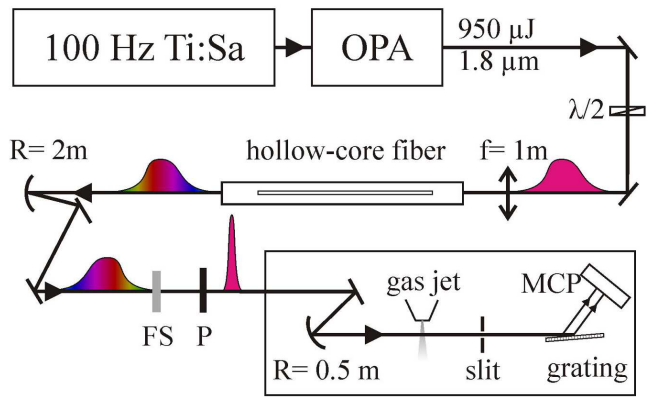}
\caption{Experimental setup.}
\label{fig:Exp_Setup}
\end{figure}

 The experimental setup as shown
in Fig. \ref{fig:Exp_Setup} starts with a Ti:Sa 800nm, 35 femtosecond laser
delivering 6 mJ of energy per pulse, at a repetition rate of 100Hz. Using the
idler beam of a fluorescence-seeded commercial optical parametric amplifier (OPA, HE TOPAS from
Light Conversion) the 800nm pulses are converted to 1825 nm. The idler signal
from the OPA produces pulses with 52 fs duration and 900 $\mu J$ which are then
focused into a 400 $\mu m$ hollow core fiber. Spectral broadening occurs in the
capillary due to self-phase modulation and the pulses are compressed by linear
propagation through bulk material \cite{Bruno:Iddler_Compression}. The pulse duration, measured with a second harmonic generation frequency resolved optical gating
device yields 14 fs, corresponding roughly to a 2 cycle pulse. Following the fiber, the beam
is collimated with a 1m lens and then focused in a thin (500 $\mu m$) pulsed
jet of atoms or molecules. Before entering the chamber the beam passes through
a half wave plate and a nanoparticle thin film polarizer (Thorlabs LPMIR) that
allows for intensity control while keeping all other optical parameters
constant. Intensity calibration is done using the method outlined in
\cite{Shiner:wavelength_scaling}. In short, we measure the ions in parallel
with the harmonic emission as a function of laser energy and  then fitted to a
space and time averaged Yudin-Ivanov ionization model
\cite{Yudin-Ivanov_Ionization} in a cylindrical geometry
\cite{intensity_calibration}. The resulting equations are only intensity
dependent once the pulse duration and wavelength are known. All other pulse and
detector parameters enter as a single scaling coefficient. Our XUV spectrometer uses a 1200 lines/mm ruled grating (Hitachi), which disperses the light into a microchannel plate/phosphor screen detector (Burle APD 3115 32/25/8 I EDR MgF2 P20). An estimate of the spectrometer resolution can be made by looking at the recombination lines of ionized $O_2$ using circularly polarized light \cite{Gur:HHG_Calibration}. Through this method we estimate an spectral resolution of $\leq 1eV$ in the range of 30eV to 70eV.

%

\begin{figure}
\includegraphics[scale=0.6]{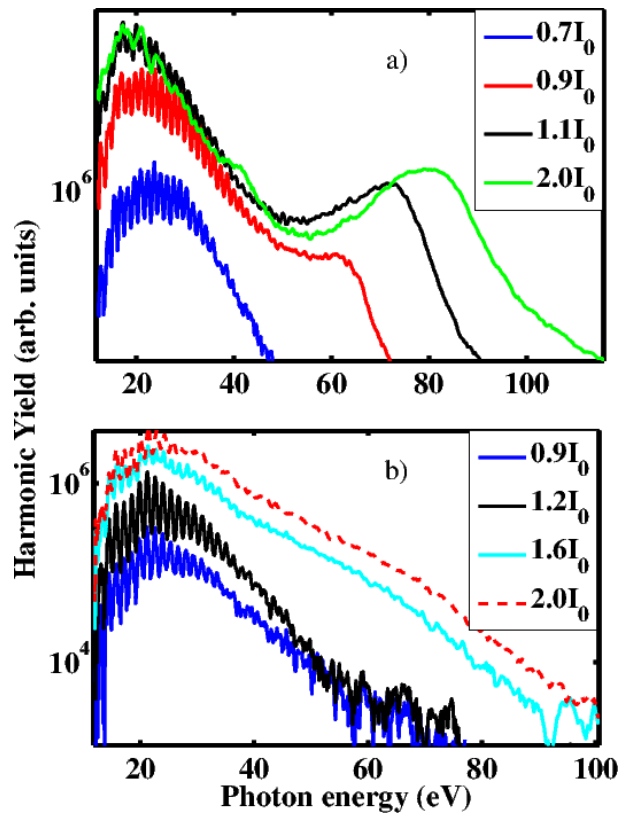}
\caption{Experimentally measured HHG spectra in Xe (a) and NO (b) for different
laser intensities. Driving field has $\lambda_L=1825nm$ and duration of 14 fs. Intensity values are in units of $I_0=10^{14}W/cm^2$}
\label{fig:Exp_Xe}
\end{figure}

In Fig.~\ref{fig:Exp_Xe} we show the HHG spectra obtained in atomic Xe and
molecular NO at different laser intensities. The most striking feature is that
the harmonic spectra becomes a near-continuum as laser intensity increases.
This effect is particularly strong for low photon energies ($\Omega < 40$ eV).
These continuous spectra extend over a broad range of photon energy from the
cutoff at about 100 eV down to very low energy of 20 - 30 eV. We also observe
this feature in other atomic targets as well as in other molecules such as
N$_2$, O$_2$, CO, CO$_2$, and other organic molecules with relatively low
ionization potentials. We therefore believe that this behavior is quite
universal and is independent of atomic and/or molecule species. We also note a
saturation effect in observed HHG spectra, when further increase of laser
intensity does not change HHG spectrum significantly. This can be clearly seen
in case of Xe for laser intensities above about $2\times 10^{14}$ W/cm$^2$ in
Fig.~1(a).

As the appearance of a broad continuum is an indication of a possible IAP, it
is very tempting to speculate this transition as a transition from an APT to
IAP by controlling the intensity of few-cycle mid-IR pulses. The transition from APT to IAP at very high laser intensities has has been demonstrated previously where the mechanism was ionization gating \cite{Leone:Ionization_gating_HHG,
nisoli:high_energy_atto}. In ionization gating, the field is so intense that
the electron is momentarily stripped from the atomic core, thus making the
electric field effectively one-cycle. To confirm our hypothesis of the
transition from APT to IAP and to further understand the underlying physics we
carried out theoretical simulations, which include the macroscopic propagation
of both driving laser and HHG fields, see below. Our simulations reveal that
the mechanism behind this transition is quite different from the conventional
ionization gating. In fact, the mechanism is closer to the ionization-driven
reshaping of the driving laser pulse, as discussed by Gaarde {\it et al}
\cite{Gaarde:Atto_spatio_temporal, gaarde:ATTO_filter}. However, the use of
mid-IR laser pulse leads to a different regime, where the gas pressure is much
lower, and the gas jet is much thinner, as compared to that of Gaarde {\it et
al} \cite{Gaarde:Atto_spatio_temporal, gaarde:ATTO_filter}.

Before presenting results of our simulation we first comment on the differences between the experimental conditions for the present work and for our earlier results
\cite{Shiner:2e_Xe}.  The conditions for \cite{Shiner:2e_Xe} were chosen to avoid phase mismatch so as  to measure harmonic spectra that come as close as possible to the single atom response.  In doing so a broad enhancement in HHG yield was observed near 100 eV, which is attributed to the effect of the interchannel coupling in the photoioinzatin / photo-recombination cross section of Xe \cite{Shiner:2e_Xe}.  To access the single atom response the laser was focused 1-2mm downstream from the jet where the gas density was relatively low supporting harmonics which were well phase matched across the entire xuv spectrum.  In the present work, the focus was brought closer to the jet into a region where the gas density was higher.  This lead to the sub-cycle modification of the laser field that we discuss in the following section and also explains why the cutoff observed after macroscopic propagation falls short of the cutoff expected based on the laser intensity and the single atom response.  The laser setup used for the present work is identical to that in \cite{Shiner:2e_Xe} although the pulse duration is somewhat longer (14 fs vs 11fs).  The small bump near 40eV for the highest intensity of $2.0 \times 10^{14}$ W/cm$^2$ in Fig.\ref{fig:Exp_Xe}a) is an artifact of the grating used for the present study.  To explain the spectrum that we observe in the present work we  performed simulations in atomic Xe at different focusing positions, gas pressure, and pulse durations. We found that the most influential parameters are the pressure and the size of the gas jet

\begin{figure}
\centering
\includegraphics[scale=0.65, angle=-90]{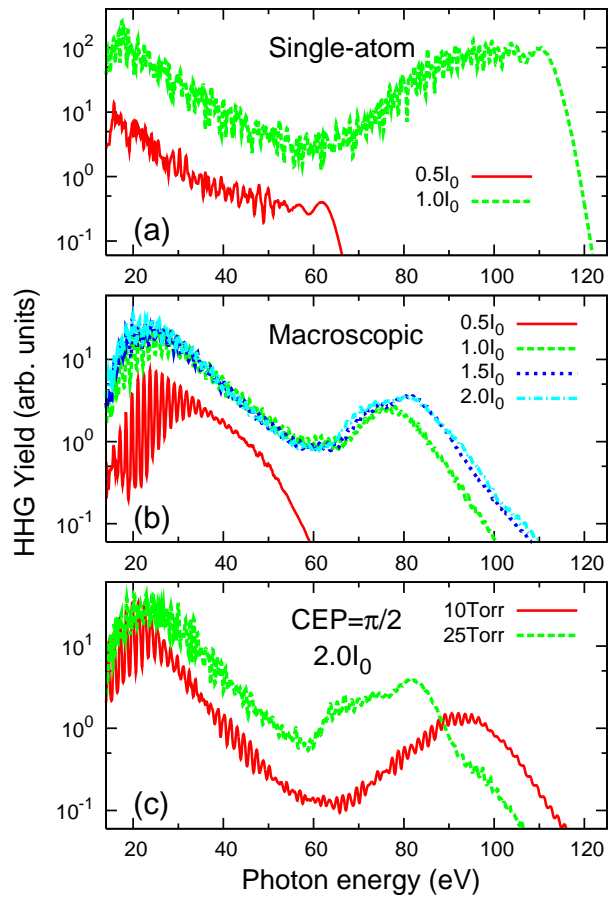}
\caption{(color online) HHG spectra calculated using QRS theory for different
peak intensities as indicated where $I_{0}$=10$^{14}$ W/cm$^2$. Panel (a) shows
spectra for a single atom response. Panel (b) shows CEP-averaged spectra from
macroscopic simulations for different intensities. Panel (c) shows spectra from
macroscopic simulations for gas pressures of 10 and 25 Torr, at a CEP of
$\pi/2$. Other laser parameters are the same as in the experiment for all
panels.} \label{fig:HHG_Theory}
\end{figure}

For an in-depth detail of the theoretical method used in this paper, we point
the reader towards Ref. \cite{Jin:Xe_Atto_HHG}. We only briefly describe the
method here. The theoretical study of high harmonic generation (HHG) consists
of two parts \cite{Jin:PRA2009,Jin:PRA2011,Jin:JPB2011}. The first one is to
calculate the induced dipole of each atom or molecule in the laser field. The second part is to consider the nonlinear propagation of the
fundamental laser pulse and the harmonic fields in the medium by solving the
Maxwell's equations. For the propagation of the fundamental laser field, we
include the effects of refraction, nonlinear Kerr effect, ionization, and
plasma defocusing \cite{Jin:PRA2011}. Due to the absence of such parameters for most molecules, our simulations focus mainly on atomic Xe. We use the recently developed
quantitative rescattering (QRS) theory
\cite{CDLin:JPB-Review,Toru:PRL2008,ATLe:QRS2009} to calculate the induced
dipole of single atom or molecule. Once the induced dipoles are calculated for
different laser intensities in the medium, they are then fed into the
propagation equations of harmonic field. At the exit face of the medium, we
obtain the near-field harmonic emission. Harmonics then propagate further in
the free space, and they may go through a slit before they are collected by the
detector. The far-field harmonic emission can be obtained from the near-field
harmonic emissions through a Hankel transformation. We pay particular attention
to the multi-electron correlated process that occurs in atomic Xe for returning
electrons with energies beyond ~90 eV. The QRS states that the single-atom
response induced dipole $D(\omega)$ can be written as
\begin{equation}
D(\omega) = W(\omega) d(\omega),
\end{equation}
with $W(\omega)$ the laser-driven macroscopic wave packet and $d(\omega)$ the
photo-recombination (PR) dipole moment. To describe the PR process of Xe, we
need to use the PR transition dipole moment including multielectron effects in
QRS theory. Such PR moment can be extracted from the experimental
photoionization cross section (PICS) \cite{Becker:PICS,Fahlman:PICS}. In the
calculation, we incorporate the PICS of Xe, which is extended into high-photon
energy region (a few hundred eV), using the relativistic random-phase
approximation (RRPA) by Kutzner et al. \cite{Kutzner:PICS}. Note that the phase
of PR transition dipole moment is assumed the same as that solely caused by 5p.

The results of our simulations in atomic Xe are shown in
Fig.~\ref{fig:HHG_Theory}. First, the CEP-averaged HHG spectra of the single
atom response obtained with the QRS method are shown in Fig.~\ref{fig:HHG_Theory}(a) for two laser
intensities of $0.5 \times 10^{14}$ W/cm$^2$ and $1.0 \times 10^{14}$ W/cm$^2$.
Clearly, the simulation at the single atom level does not reproduce the the
transition to the continuum spectrum, observed in the experiment, as well as
the HHG cutoff position for the high laser intensity. The ionization probability is about 10-30\% at the
end of the pulse in this range of laser intensity ($I_L \approx 1.0 \times 10^{14}$ W/cm$^2$). This result basically
excludes the possibility of ionization gating at the single atom level, as a
mechanism for the transition to the near-continuum spectra. We also found that
further increase in laser intensity up to $2.0 \times 10^{14}$ W/cm$^2$ does
not change this behaviour.We comment that the
generation of an IAP with a high pulse energy (few nanojoules) has been
reported recently by Ferrari {\it et al} \cite{nisoli:high_energy_atto}, which
is based on complete depletion of the neutral Xe atoms.

Next, we show in Fig.~\ref{fig:HHG_Theory}(b) the harmonic spectra for four peak laser intensities
after propagation through the gas jet with parameters that closely resemble the
experimental conditions. Such condition include not only the laser parameters
but also the jet size (1 mm), the spectrometer slit opening (190 $\mu m$) and
distance of the slit to the interacting region (455 mm). In the theoretical
calculations the beam is assumed to be Gaussian. All spectra are averaged over
random values of the CEP as in the
experiment. The macroscopic simulations indeed show a transition to a
near-continuum HHG spectrum for a broad range of photon energy from the
``apparent'' cutoff down to about 30 eV, as the laser intensity increases, thus
reproducing nicely the experimental finding. The simulations also reproduce the
saturation effect observed in the experiments. This is not the case for the
single atom simulations shown in Fig.~\ref{fig:HHG_Theory}(a). This lead us to conclude that the
near-continuum is a consequence of the propagation in the medium of both the
fundamental and the harmonics fields.

By comparing HHG spectra obtained for the single atom with the ones from a
macroscopic simulation for the same peak intensities, we observe large
discrepancies in the cutoff and the overall shape. Before explaining these
discrepancies as well as the physical mechanism behind the experimentally
observed spectra we address here the differences between the current
experimental data and that of Shiner {\it et al} \cite{Shiner:2e_Xe}. Fig.~3(c)
shows a comparison between the HHG spectra from macroscopic simulations for two
different gas pressures of 10 and 20 Torr. The laser intensity is $2.0 \times
10^{14}$ W/cm$^2$ and the CEP is chosen to be $\pi/2$. The ``apparent" cut-offs
are different by about 15 eV. Thus the discrepancy between the two experiments
can be tentatively attributed as due to the different gas pressures. Moreover,
we note that the size of the gas jet could also affect the cut-off position and
the overall shape of HHG spectra. A thinner gas jet would give a spectrum
closer to a single-atom response, i.e., with a higher cut-off.

\begin{figure}
\centering
\includegraphics[angle=-90, scale = 0.6]{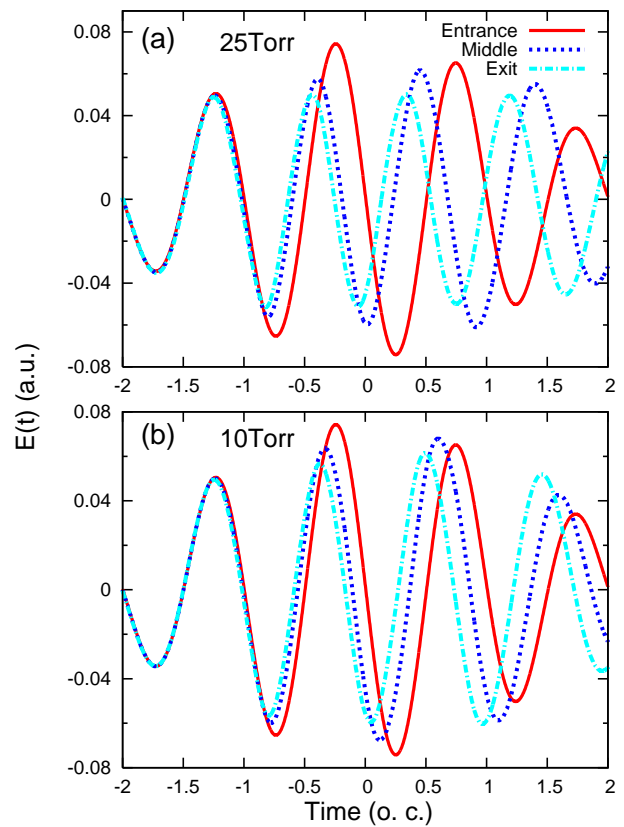}
\caption{Driving electric field ($\lambda_L = 1825nm$) as a function of time
(in units of the optical cycle) at three different positions relative to the
atomic jet, for a gas pressure of 25 Torr (a) and 10 Torr (b). All values are
``measured'' on axis. Laser parameters are $I_L=2.0 \times 10^{14}$ W/cm$^2$
and CEP=$\pi/2$.} \label{fig:Fundamental_chirp}
\end{figure}

In order to understand the mechanism behind the transition to near-continuum
HHG spectrum, we followed the fundamental field in space and time through the
simulation. In Fig.~\ref{fig:Fundamental_chirp} we show the on-axis electric
field amplitude as a function of time at the entrance, middle, and just before
the exit of the Xe gas jet, for a gas pressure of 25 Torr and 10 Torr. The CEP
is chosen to be $\pi/2$, and the laser intensity is $2.0 \times 10^{14}$
W/cm$^2$. As can be seen from the figure, the temporal shape of the laser pulse
is strongly modified as it propagates though the medium. In particular, a
dynamic blueshift (or chirp) can also be seen clearly. In fact, this shape is
quite typical for the self-phase modulation found in a rapidly ionizing medium
\cite{Gaarde:Atto_spatio_temporal, gaarde:ATTO_filter, Jin:Xe_Atto_HHG}. A
chirped pulse means that harmonics will be produced at different fundamental
frequencies depending on where and when they are generated. The effect of chirp
in HHG has already been established \cite{KM:HHG_Chirp, Nam:HHG_Chirp} and it
induces a frequency shift in the observed harmonic spectrum. In our case, the
harmonics are produced with different chirped pulses all at once. Furthermore,
the electron density strongly varies as a function of radial distance due to
the radial variation of the laser intensity. This radial variation electron
density  acts like a negative lens to defocus the laser beam. This has also
been found in the simulation (see, \cite{Jin:Xe_Atto_HHG}). We found that both
temporal and spatial reshapings are quite negligible at the laser intensity of
$0.5 \times 10^{14}$ W/cm$^2$, since the ionization is insignificant. This
explains the observed transition to the near-continuum in HHG spectra as laser
intensity increases.

The temporal and spatial reshaping of the laser beam significantly reduces the
laser intensity in the medium, which accounts for the ``apparent'' lower energy
cut-off in the HHG spectra found in the macroscopic simulations, as compared to
the single atom simulations. Further increase in laser intensity above $2.0
\times 10^{14}$ W/cm$^2$ will leads to even stronger reshaping and
redistribution of laser intensity, which explains the saturation effect
discussed above. 

The reshaping of the laser pulse has been discussed before in the context of
the 800 nm driving pulse \cite{Gaarde:Atto_spatio_temporal,
gaarde:ATTO_filter}. The effect is even more significant for the mid-IR laser,
since the effective index of refraction due to free electron is $\eta_e \sim
-n_e\lambda_L^2$, where $n_e$ is the density of free electron and $\lambda_L$
is the wavelength of the driving laser. Therefore the reshaping of the laser
pulse occurs at much lower gas pressure and size (25 Torr, 1 mm long) as
compared to that of Gaarde {\it et al} (135 Torr, 3 mm long). We further
comment that at the lower pressure of 10 Torr, shown in Fig.~4(b), the
reshaping is less severe as compared to 25 Torr, thus leads to the higher HHG
cutoff seen in Fig.~3(c).

\begin{figure}
\centering
\includegraphics[angle=-90, scale = 0.6]{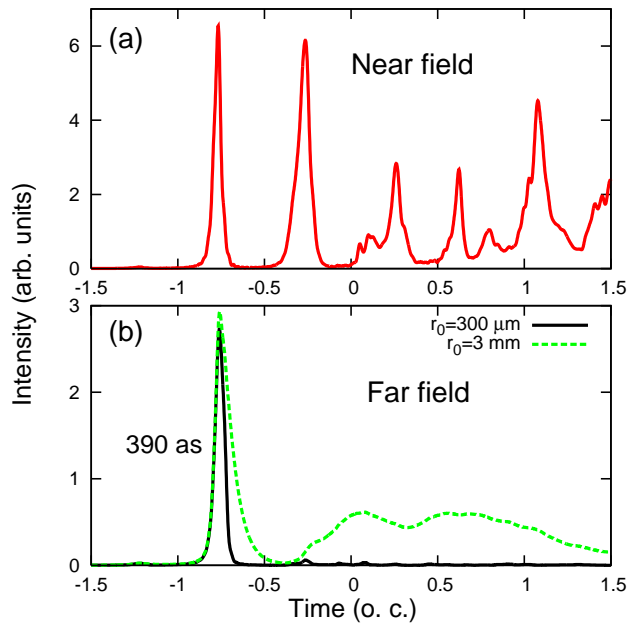}
\caption{(Color online) (a) Intensity of attosecond pulses in the near field.
(b) Intensity of isolated attosecond pulse (IAP) in the far field (measured 455 mm after
the laser focus). Black is using a spatial filter with a radius $r_o = 300 \mu m$, located 455mm after the focus. In green, a spatial filter of $r_o = 3 mm$, at the same position is used instead.
Harmonics used to synthesize attosecond pulses are H40 (27.2 eV) to H80 (54.4
eV).  Laser intensity at the focus is $1.5 \times 10^{14}$ W/cm$^2$ and the CEP
is $\pi/2$.} \label{fig:Atto_pulses_full}
\end{figure}

To further understand the nature of the near-continuum and to address the
usefulness of our scheme for the production of IAPs we follow the attosecond
pulses in time as they emerge from the interacting region. We use a spectral
filter to select harmonics from H40 (27.2eV) to H80 (54.4 eV).
Fig.~\ref{fig:Atto_pulses_full}(a) shows the attosecond pulses using these
harmonics for the same driving field [$\lambda_L = 1825$ nm and duration (FWHM)
of 14 fs] with a peak intensity of $1.5 \times 10^{14}$ W/cm$^2$, and a CEP of
$\pi/2$ at the end of the interacting region (near field). The presence of an
APT is evident. As proposed in \cite{gaarde:ATTO_filter}, an IAP can be
obtained in the far field by spatially filtering the XUV pulses. In fact, due
to the temporal and spatial reshaping, the HHG emission at different times will
have a different divergence. Fig.~\ref{fig:Atto_pulses_full}(b) shows harmonics
H40 to H80 in time domain, under the same conditions as panel (a), but the
pulse is ``measured'' in the far field after a circular
filter with a radius of $300$ $\mu$m, located 455mm after the focus. In this case, a single 390 as pulse is
indeed measured, thus proving that the observed XUV supercontinuum can indeed
produce IAPs. This attosecond burst is generated in the leading edge of the
pulse, less than one optical cycle before the peak intensity is reached.
Clearly, the more divergent attosecond bursts, generated at later times have
been blocked by the filter. This result also indicates that the phase
mismatching due to the free electron might also play an important role, as the
near-axis contribution from the later sub-cycles is suppressed due to the phase
mismatch. We mention that the phase mismatch due to the free electron is more
severe for the long wavelength. Figure \ref{fig:Atto_pulses_full}(b) also shows in green (color online) the XUV pulse under the same conditions but using a 3mm spatial filter. In this case, an IAP is not obtained, thus showing the importance of the filtering. We also carried out simulations with a lower
laser intensity of $0.5 \times 10^{14}$ W/cm$^2$. The result shows an APT even
after the spatial filter is applied. At this low intensity, the ionization and
the pulse reshaping are insignificant. Therefore, in this case different XUV
bursts at different half-cycles will have similar divergence.

We emphasize that relatively low harmonics is chosen here to illustrate the
usefulness of our scheme for producing an IAP with plateau harmonics. In
contrast, in a typical IAP generation with the reshaping mechanism with 800 nm
laser pulses, only the harmonics near the cutoff are used for producing an IAP
\cite{gaarde:ATTO_filter}. In Ref.~\cite{Jin:Xe_Atto_HHG} we provide further
details on how this IAP can be manipulated by changing the CEP of the driving
field, as well as the spectral and spatial filters. In particular, it was shown
that the IAPs can be produced even without CEP stabilized laser pulses for most
of the CEPs. Although the experimental results and theoretical analysis
presented here are for two-cycle pulses, we anticipate that production of IAPs
within our scheme is also feasible with longer pulses.


In conclusion, we have observed a near continuum in the HHG spectrum, with a 1.8 $\mu$m laser driver as the laser intensity increases. By
performing theoretical simulations of the full macroscopic HHG process combined
with a realistic single atom response calculation we are able to reproduce the
same behavior as we observe in the experiment. The simulations show that the
transition from a train of attosecond pulses to an isolated attosecond pulse is
due to strong reshaping of the driving laser as it propagates through the
medium. The main advantage of our approach for producing IAPs is that the
central frequency of the XUV pulse can be tuned over a broader range of photon
energy as compared to the standard 800 nm. This is due to the capability of
using a large range of harmonics, including plateau harmonics, of a broad
near-continuum spectrum. We comment that our method is based on a different
mechanism from that of Ferrari {\it et al} \cite{nisoli:high_energy_atto},
which is based on complete depletion of the neutral atoms. However, in both
methods, both plateau and cutoff harmonics can be used to synthesize IAPs.
Finally, it has been reported recently \cite{KM:Phase_Matching} that the XUV
yield can increase with wavelength when a proper phase matching conditions are
exploited, thus creating a great venue for our approach.

\section*{Acknowledments}
This work was supported in part by the Chemical Sciences, Geosciences and
Biosciences Division, Office of Basic Energy Sciences, Office of Science, U. S.
Department of Energy. CT-H also thanks the AFOSR for their financial support.

\bibliography{carlos_bib_v3}

\begin{thebibliography}{44}
\expandafter\ifx\csname natexlab\endcsname\relax\def\natexlab#1{#1}\fi
\expandafter\ifx\csname bibnamefont\endcsname\relax
  \def\bibnamefont#1{#1}\fi
\expandafter\ifx\csname bibfnamefont\endcsname\relax
  \def\bibfnamefont#1{#1}\fi
\expandafter\ifx\csname citenamefont\endcsname\relax
  \def\citenamefont#1{#1}\fi
\expandafter\ifx\csname url\endcsname\relax
  \def\url#1{\texttt{#1}}\fi
\expandafter\ifx\csname urlprefix\endcsname\relax\def\urlprefix{URL }\fi
\providecommand{\bibinfo}[2]{#2}
\providecommand{\eprint}[2][]{\url{#2}}

\bibitem[{\citenamefont{Hentschel et~al.}(2001)\citenamefont{Hentschel,
  Kienberger, Spielmann, Reider, Milosevic, Brabec, Corkum, Heinzmann,
  Drescher, and Krausz}}]{Krausz:Attosecond_Metrology}
\bibinfo{author}{\bibfnamefont{M.}~\bibnamefont{Hentschel}},
  \bibinfo{author}{\bibfnamefont{R.}~\bibnamefont{Kienberger}},
  \bibinfo{author}{\bibfnamefont{C.}~\bibnamefont{Spielmann}},
  \bibinfo{author}{\bibfnamefont{G.~A.} \bibnamefont{Reider}},
  \bibinfo{author}{\bibfnamefont{N.}~\bibnamefont{Milosevic}},
  \bibinfo{author}{\bibfnamefont{T.}~\bibnamefont{Brabec}},
  \bibinfo{author}{\bibfnamefont{P.}~\bibnamefont{Corkum}},
  \bibinfo{author}{\bibfnamefont{U.}~\bibnamefont{Heinzmann}},
  \bibinfo{author}{\bibfnamefont{M.}~\bibnamefont{Drescher}}, \bibnamefont{and}
  \bibinfo{author}{\bibfnamefont{F.}~\bibnamefont{Krausz}},
  \bibinfo{journal}{Nature} \textbf{\bibinfo{volume}{414}},
  \bibinfo{pages}{509} (\bibinfo{year}{2001}).

\bibitem[{\citenamefont{Paul et~al.}(2001)\citenamefont{Paul, Toma, Breger,
  Mullot, Aug\'e, Balcou, Muller, and Agostini}}]{Agostini:Attosecond_train}
\bibinfo{author}{\bibfnamefont{P.~M.} \bibnamefont{Paul}},
  \bibinfo{author}{\bibfnamefont{E.~S.} \bibnamefont{Toma}},
  \bibinfo{author}{\bibfnamefont{P.}~\bibnamefont{Breger}},
  \bibinfo{author}{\bibfnamefont{G.}~\bibnamefont{Mullot}},
  \bibinfo{author}{\bibfnamefont{F.}~\bibnamefont{Aug\'e}},
  \bibinfo{author}{\bibfnamefont{P.}~\bibnamefont{Balcou}},
  \bibinfo{author}{\bibfnamefont{H.~G.} \bibnamefont{Muller}},
  \bibnamefont{and} \bibinfo{author}{\bibfnamefont{P.}~\bibnamefont{Agostini}},
  \bibinfo{journal}{Science} \textbf{\bibinfo{volume}{292}},
  \bibinfo{pages}{1689} (\bibinfo{year}{2001}).

\bibitem[{\citenamefont{Corkum and Krausz}(2007)}]{corkum:attosecond_review}
\bibinfo{author}{\bibfnamefont{P.~B.} \bibnamefont{Corkum}} \bibnamefont{and}
  \bibinfo{author}{\bibfnamefont{F.}~\bibnamefont{Krausz}},
  \bibinfo{journal}{Nat. Phys.} \textbf{\bibinfo{volume}{3}},
  \bibinfo{pages}{381} (\bibinfo{year}{2007}).

\bibitem[{\citenamefont{Krausz and
  Ivanov}(2009)}]{Krausz_Ivanov:Attosecond_review}
\bibinfo{author}{\bibfnamefont{F.}~\bibnamefont{Krausz}} \bibnamefont{and}
  \bibinfo{author}{\bibfnamefont{M.}~\bibnamefont{Ivanov}},
  \bibinfo{journal}{Rev. Mod. Phys.} \textbf{\bibinfo{volume}{81}},
  \bibinfo{pages}{163} (\bibinfo{year}{2009}).

\bibitem[{\citenamefont{Corkum and Chang}(2008)}]{Corkum_Chang:Attosecond}
\bibinfo{author}{\bibfnamefont{P.~B.} \bibnamefont{Corkum}} \bibnamefont{and}
  \bibinfo{author}{\bibfnamefont{Z.}~\bibnamefont{Chang}},
  \bibinfo{journal}{Opt. Photon. News} \textbf{\bibinfo{volume}{19}},
  \bibinfo{pages}{24} (\bibinfo{year}{2008}).

\bibitem[{\citenamefont{Christov et~al.}(1997)\citenamefont{Christov, Murnane,
  and Kapteyn}}]{KM:Attosecond_Pulses_Theory_1997}
\bibinfo{author}{\bibfnamefont{I.~P.} \bibnamefont{Christov}},
  \bibinfo{author}{\bibfnamefont{M.~M.} \bibnamefont{Murnane}},
  \bibnamefont{and} \bibinfo{author}{\bibfnamefont{H.~C.}
  \bibnamefont{Kapteyn}}, \bibinfo{journal}{Phys. Rev. Lett.}
  \textbf{\bibinfo{volume}{78}}, \bibinfo{pages}{1251} (\bibinfo{year}{1997}).

\bibitem[{\citenamefont{Popmintchev et~al.}(2010)\citenamefont{Popmintchev,
  Chen, Arpin, Murnane, and Kapteyn}}]{JILA:review2010}
\bibinfo{author}{\bibfnamefont{T.}~\bibnamefont{Popmintchev}},
  \bibinfo{author}{\bibfnamefont{M.~C.} \bibnamefont{Chen}},
  \bibinfo{author}{\bibfnamefont{P.}~\bibnamefont{Arpin}},
  \bibinfo{author}{\bibfnamefont{M.~M.} \bibnamefont{Murnane}},
  \bibnamefont{and} \bibinfo{author}{\bibfnamefont{H.~C.}
  \bibnamefont{Kapteyn}}, \bibinfo{journal}{Nat. Photon.}
  \textbf{\bibinfo{volume}{4}}, \bibinfo{pages}{822} (\bibinfo{year}{2010}).

\bibitem[{\citenamefont{Goulielmakis et~al.}(2008)\citenamefont{Goulielmakis,
  Schultze, Hofstetter, Yakovlev, Gagnon, Uiberacker, Aquila, Gullikson,
  Attwood, Kienberger et~al.}}]{Goulielmakis:Single_Attosecond}
\bibinfo{author}{\bibfnamefont{E.}~\bibnamefont{Goulielmakis}},
  \bibinfo{author}{\bibfnamefont{M.}~\bibnamefont{Schultze}},
  \bibinfo{author}{\bibfnamefont{M.}~\bibnamefont{Hofstetter}},
  \bibinfo{author}{\bibfnamefont{V.~S.} \bibnamefont{Yakovlev}},
  \bibinfo{author}{\bibfnamefont{J.}~\bibnamefont{Gagnon}},
  \bibinfo{author}{\bibfnamefont{M.}~\bibnamefont{Uiberacker}},
  \bibinfo{author}{\bibfnamefont{A.~L.} \bibnamefont{Aquila}},
  \bibinfo{author}{\bibfnamefont{E.~M.} \bibnamefont{Gullikson}},
  \bibinfo{author}{\bibfnamefont{D.~T.} \bibnamefont{Attwood}},
  \bibinfo{author}{\bibfnamefont{R.}~\bibnamefont{Kienberger}},
  \bibnamefont{et~al.}, \bibinfo{journal}{Science}
  \textbf{\bibinfo{volume}{320}}, \bibinfo{pages}{1614} (\bibinfo{year}{2008}).

\bibitem[{\citenamefont{Corkum et~al.}(1994)\citenamefont{Corkum, Burnett, and
  Ivanov}}]{Corkum:polarization_gating}
\bibinfo{author}{\bibfnamefont{P.~B.} \bibnamefont{Corkum}},
  \bibinfo{author}{\bibfnamefont{N.~H.} \bibnamefont{Burnett}},
  \bibnamefont{and} \bibinfo{author}{\bibfnamefont{M.~Y.}
  \bibnamefont{Ivanov}}, \bibinfo{journal}{Opt. Lett.}
  \textbf{\bibinfo{volume}{19}}, \bibinfo{pages}{1870} (\bibinfo{year}{1994}).

\bibitem[{\citenamefont{Tcherbakoff et~al.}(2003)\citenamefont{Tcherbakoff,
  M\'evel, Descamps, Plumridge, and
  Constant}}]{Constant:attosecond_polarization_gating}
\bibinfo{author}{\bibfnamefont{O.}~\bibnamefont{Tcherbakoff}},
  \bibinfo{author}{\bibfnamefont{E.}~\bibnamefont{M\'evel}},
  \bibinfo{author}{\bibfnamefont{D.}~\bibnamefont{Descamps}},
  \bibinfo{author}{\bibfnamefont{J.}~\bibnamefont{Plumridge}},
  \bibnamefont{and} \bibinfo{author}{\bibfnamefont{E.}~\bibnamefont{Constant}},
  \bibinfo{journal}{Phys. Rev. A} \textbf{\bibinfo{volume}{68}},
  \bibinfo{pages}{043804} (\bibinfo{year}{2003}).

\bibitem[{\citenamefont{Sola et~al.}(2006)\citenamefont{Sola, Mevel, Elouga,
  Constant, Strelkov, Poletto, Villoresi, Benedetti, Caumes, Stagira
  et~al.}}]{nisoli:polarization_gating}
\bibinfo{author}{\bibfnamefont{I.~J.} \bibnamefont{Sola}},
  \bibinfo{author}{\bibfnamefont{E.}~\bibnamefont{Mevel}},
  \bibinfo{author}{\bibfnamefont{L.}~\bibnamefont{Elouga}},
  \bibinfo{author}{\bibfnamefont{E.}~\bibnamefont{Constant}},
  \bibinfo{author}{\bibfnamefont{V.}~\bibnamefont{Strelkov}},
  \bibinfo{author}{\bibfnamefont{L.}~\bibnamefont{Poletto}},
  \bibinfo{author}{\bibfnamefont{P.}~\bibnamefont{Villoresi}},
  \bibinfo{author}{\bibfnamefont{E.}~\bibnamefont{Benedetti}},
  \bibinfo{author}{\bibfnamefont{J.-P.} \bibnamefont{Caumes}},
  \bibinfo{author}{\bibfnamefont{S.}~\bibnamefont{Stagira}},
  \bibnamefont{et~al.}, \bibinfo{journal}{Nat. Phys.}
  \textbf{\bibinfo{volume}{2}}, \bibinfo{pages}{319} (\bibinfo{year}{2006}).

\bibitem[{\citenamefont{Mashiko et~al.}(2008)\citenamefont{Mashiko, Gilbertson,
  Li, Khan, Shakya, Moon, and Chang}}]{Chang:double_optical_gating}
\bibinfo{author}{\bibfnamefont{H.}~\bibnamefont{Mashiko}},
  \bibinfo{author}{\bibfnamefont{S.}~\bibnamefont{Gilbertson}},
  \bibinfo{author}{\bibfnamefont{C.}~\bibnamefont{Li}},
  \bibinfo{author}{\bibfnamefont{S.~D.} \bibnamefont{Khan}},
  \bibinfo{author}{\bibfnamefont{M.~M.} \bibnamefont{Shakya}},
  \bibinfo{author}{\bibfnamefont{E.}~\bibnamefont{Moon}}, \bibnamefont{and}
  \bibinfo{author}{\bibfnamefont{Z.}~\bibnamefont{Chang}},
  \bibinfo{journal}{Phys. Rev. Lett.} \textbf{\bibinfo{volume}{100}},
  \bibinfo{pages}{103906} (\bibinfo{year}{2008}).

\bibitem[{\citenamefont{Feng et~al.}(2009)\citenamefont{Feng, Gilbertson,
  Mashiko, Wang, Khan, Chini, Wu, Zhao, and Chang}}]{Chang:GDOG}
\bibinfo{author}{\bibfnamefont{X.}~\bibnamefont{Feng}},
  \bibinfo{author}{\bibfnamefont{S.}~\bibnamefont{Gilbertson}},
  \bibinfo{author}{\bibfnamefont{H.}~\bibnamefont{Mashiko}},
  \bibinfo{author}{\bibfnamefont{H.}~\bibnamefont{Wang}},
  \bibinfo{author}{\bibfnamefont{S.~D.} \bibnamefont{Khan}},
  \bibinfo{author}{\bibfnamefont{M.}~\bibnamefont{Chini}},
  \bibinfo{author}{\bibfnamefont{Y.}~\bibnamefont{Wu}},
  \bibinfo{author}{\bibfnamefont{K.}~\bibnamefont{Zhao}}, \bibnamefont{and}
  \bibinfo{author}{\bibfnamefont{Z.}~\bibnamefont{Chang}},
  \bibinfo{journal}{Phys. Rev. Lett.} \textbf{\bibinfo{volume}{103}},
  \bibinfo{pages}{183901} (\bibinfo{year}{2009}).

\bibitem[{\citenamefont{Gilbertson et~al.}(2010)\citenamefont{Gilbertson, Khan,
  Wu, Chini, and Chang}}]{Zhang:GDOG_no_CEP}
\bibinfo{author}{\bibfnamefont{S.}~\bibnamefont{Gilbertson}},
  \bibinfo{author}{\bibfnamefont{S.~D.} \bibnamefont{Khan}},
  \bibinfo{author}{\bibfnamefont{Y.}~\bibnamefont{Wu}},
  \bibinfo{author}{\bibfnamefont{M.}~\bibnamefont{Chini}}, \bibnamefont{and}
  \bibinfo{author}{\bibfnamefont{Z.}~\bibnamefont{Chang}},
  \bibinfo{journal}{Phys. Rev. Lett.} \textbf{\bibinfo{volume}{105}},
  \bibinfo{pages}{093902} (\bibinfo{year}{2010}).

\bibitem[{\citenamefont{Gaarde et~al.}(2008)\citenamefont{Gaarde, Tate, and
  Schafer}}]{gaarde:ATTO_filter}
\bibinfo{author}{\bibfnamefont{M.~B.} \bibnamefont{Gaarde}},
  \bibinfo{author}{\bibfnamefont{J.~L.} \bibnamefont{Tate}}, \bibnamefont{and}
  \bibinfo{author}{\bibfnamefont{K.~J.} \bibnamefont{Schafer}},
  \bibinfo{journal}{J. Phys. B: At. Mol. Opt. Phys.}
  \textbf{\bibinfo{volume}{41}}, \bibinfo{pages}{132001}
  (\bibinfo{year}{2008}).

\bibitem[{\citenamefont{Pfeifer et~al.}(2007)\citenamefont{Pfeifer, Jullien,
  Abel, Nagel, Gallmann, Neumark, and Leone}}]{Pfeifer_Leone:Ionization_gating}
\bibinfo{author}{\bibfnamefont{T.}~\bibnamefont{Pfeifer}},
  \bibinfo{author}{\bibfnamefont{A.}~\bibnamefont{Jullien}},
  \bibinfo{author}{\bibfnamefont{M.~J.} \bibnamefont{Abel}},
  \bibinfo{author}{\bibfnamefont{P.~M.} \bibnamefont{Nagel}},
  \bibinfo{author}{\bibfnamefont{L.}~\bibnamefont{Gallmann}},
  \bibinfo{author}{\bibfnamefont{D.~M.} \bibnamefont{Neumark}},
  \bibnamefont{and} \bibinfo{author}{\bibfnamefont{S.~R.} \bibnamefont{Leone}},
  \bibinfo{journal}{Opt. Express} \textbf{\bibinfo{volume}{15}},
  \bibinfo{pages}{17120} (\bibinfo{year}{2007}).

\bibitem[{\citenamefont{Abel et~al.}(2009)\citenamefont{Abel, Pfeifer, Nagel,
  Boutu, Bell, Steiner, Neumark, and Leone}}]{Leone:Ionization_gating_HHG}
\bibinfo{author}{\bibfnamefont{M.~J.} \bibnamefont{Abel}},
  \bibinfo{author}{\bibfnamefont{T.}~\bibnamefont{Pfeifer}},
  \bibinfo{author}{\bibfnamefont{P.~M.} \bibnamefont{Nagel}},
  \bibinfo{author}{\bibfnamefont{W.}~\bibnamefont{Boutu}},
  \bibinfo{author}{\bibfnamefont{M.~J.} \bibnamefont{Bell}},
  \bibinfo{author}{\bibfnamefont{C.~P.} \bibnamefont{Steiner}},
  \bibinfo{author}{\bibfnamefont{D.~M.} \bibnamefont{Neumark}},
  \bibnamefont{and} \bibinfo{author}{\bibfnamefont{S.~R.} \bibnamefont{Leone}},
  \bibinfo{journal}{Chemical Physics} \textbf{\bibinfo{volume}{366}},
  \bibinfo{pages}{9 } (\bibinfo{year}{2009}).

\bibitem[{\citenamefont{Thomann et~al.}(2009)\citenamefont{Thomann, Bahabad,
  Liu, Trebino, Murnane, and Kapteyn}}]{Thomann:2009}
\bibinfo{author}{\bibfnamefont{I.}~\bibnamefont{Thomann}},
  \bibinfo{author}{\bibfnamefont{A.}~\bibnamefont{Bahabad}},
  \bibinfo{author}{\bibfnamefont{X.}~\bibnamefont{Liu}},
  \bibinfo{author}{\bibfnamefont{R.}~\bibnamefont{Trebino}},
  \bibinfo{author}{\bibfnamefont{M.~M.} \bibnamefont{Murnane}},
  \bibnamefont{and} \bibinfo{author}{\bibfnamefont{H.~C.}
  \bibnamefont{Kapteyn}}, \bibinfo{journal}{Opt. Express}
  \textbf{\bibinfo{volume}{17}}, \bibinfo{pages}{4611} (\bibinfo{year}{2009}).

\bibitem[{\citenamefont{Haworth et~al.}(2007)\citenamefont{Haworth,
  Chipperfield, Robinson, Knight, Marangos, and Tisch}}]{Marangos:HCO_2007}
\bibinfo{author}{\bibfnamefont{C.~A.} \bibnamefont{Haworth}},
  \bibinfo{author}{\bibfnamefont{L.~E.} \bibnamefont{Chipperfield}},
  \bibinfo{author}{\bibfnamefont{J.~S.} \bibnamefont{Robinson}},
  \bibinfo{author}{\bibfnamefont{P.~L.} \bibnamefont{Knight}},
  \bibinfo{author}{\bibfnamefont{J.~P.} \bibnamefont{Marangos}},
  \bibnamefont{and} \bibinfo{author}{\bibfnamefont{J.~W.~G.}
  \bibnamefont{Tisch}}, \bibinfo{journal}{Nat. Phys.}
  \textbf{\bibinfo{volume}{3}}, \bibinfo{pages}{52} (\bibinfo{year}{2007}).

\bibitem[{\citenamefont{Gaarde and
  Schafer}(2006)}]{Gaarde:Atto_spatio_temporal}
\bibinfo{author}{\bibfnamefont{M.~B.} \bibnamefont{Gaarde}} \bibnamefont{and}
  \bibinfo{author}{\bibfnamefont{K.~J.} \bibnamefont{Schafer}},
  \bibinfo{journal}{Opt. Lett.} \textbf{\bibinfo{volume}{31}},
  \bibinfo{pages}{3188} (\bibinfo{year}{2006}).

\bibitem[{\citenamefont{Takahashi et~al.}(2010)\citenamefont{Takahashi, Lan,
  M\"ucke, Nabekawa, and Midorikawa}}]{Midorikawa:2color_2010}
\bibinfo{author}{\bibfnamefont{E.~J.} \bibnamefont{Takahashi}},
  \bibinfo{author}{\bibfnamefont{P.}~\bibnamefont{Lan}},
  \bibinfo{author}{\bibfnamefont{O.~D.} \bibnamefont{M\"ucke}},
  \bibinfo{author}{\bibfnamefont{Y.}~\bibnamefont{Nabekawa}}, \bibnamefont{and}
  \bibinfo{author}{\bibfnamefont{K.}~\bibnamefont{Midorikawa}},
  \bibinfo{journal}{Phys. Rev. Lett.} \textbf{\bibinfo{volume}{104}},
  \bibinfo{pages}{233901} (\bibinfo{year}{2010}).

\bibitem[{\citenamefont{Tate et~al.}(2007)\citenamefont{Tate, Auguste, Muller,
  Sali\`eres, Agostini, and DiMauro}}]{Tate:Wavelength_scaling}
\bibinfo{author}{\bibfnamefont{J.}~\bibnamefont{Tate}},
  \bibinfo{author}{\bibfnamefont{T.}~\bibnamefont{Auguste}},
  \bibinfo{author}{\bibfnamefont{H.~G.} \bibnamefont{Muller}},
  \bibinfo{author}{\bibfnamefont{P.}~\bibnamefont{Sali\`eres}},
  \bibinfo{author}{\bibfnamefont{P.}~\bibnamefont{Agostini}}, \bibnamefont{and}
  \bibinfo{author}{\bibfnamefont{L.~F.} \bibnamefont{DiMauro}},
  \bibinfo{journal}{Phys. Rev. Lett.} \textbf{\bibinfo{volume}{98}},
  \bibinfo{pages}{013901} (\bibinfo{year}{2007}).

\bibitem[{\citenamefont{Shiner et~al.}(2009)\citenamefont{Shiner,
  Trallero-Herrero, Kajumba, Bandulet, Comtois, L\'egar\'e, Kieffer, Corkum,
  and Villeneuve}}]{Shiner:wavelength_scaling}
\bibinfo{author}{\bibfnamefont{A.~D.} \bibnamefont{Shiner}},
  \bibinfo{author}{\bibfnamefont{C.}~\bibnamefont{Trallero-Herrero}},
  \bibinfo{author}{\bibfnamefont{N.}~\bibnamefont{Kajumba}},
  \bibinfo{author}{\bibfnamefont{H.-C.} \bibnamefont{Bandulet}},
  \bibinfo{author}{\bibfnamefont{D.}~\bibnamefont{Comtois}},
  \bibinfo{author}{\bibfnamefont{F.}~\bibnamefont{L\'egar\'e}},
  \bibinfo{author}{\bibfnamefont{J.-C.} \bibnamefont{Kieffer}},
  \bibinfo{author}{\bibfnamefont{P.~B.} \bibnamefont{Corkum}},
  \bibnamefont{and} \bibinfo{author}{\bibfnamefont{D.~M.}
  \bibnamefont{Villeneuve}}, \bibinfo{journal}{Phys. Rev. Lett.}
  \textbf{\bibinfo{volume}{103}}, \bibinfo{pages}{07390}
  (\bibinfo{year}{2009}).

\bibitem[{\citenamefont{Frolov et~al.}(2008)\citenamefont{Frolov, Manakov, and
  Starace}}]{Starace:wavelength_scaling}
\bibinfo{author}{\bibfnamefont{M.~V.} \bibnamefont{Frolov}},
  \bibinfo{author}{\bibfnamefont{N.~L.} \bibnamefont{Manakov}},
  \bibnamefont{and} \bibinfo{author}{\bibfnamefont{A.~F.}
  \bibnamefont{Starace}}, \bibinfo{journal}{Phys. Rev. Lett.}
  \textbf{\bibinfo{volume}{100}}, \bibinfo{pages}{173001}
  (\bibinfo{year}{2008}).

\bibitem[{\citenamefont{Schiessl et~al.}(2007)\citenamefont{Schiessl, Ishikawa,
  Persson, and Burgd\"orfer}}]{Burgdorfer:Wavelength_scaling}
\bibinfo{author}{\bibfnamefont{K.}~\bibnamefont{Schiessl}},
  \bibinfo{author}{\bibfnamefont{K.~L.} \bibnamefont{Ishikawa}},
  \bibinfo{author}{\bibfnamefont{E.}~\bibnamefont{Persson}}, \bibnamefont{and}
  \bibinfo{author}{\bibfnamefont{J.}~\bibnamefont{Burgd\"orfer}},
  \bibinfo{journal}{Phys. Rev. Lett.} \textbf{\bibinfo{volume}{99}},
  \bibinfo{pages}{253903} (\bibinfo{year}{2007}).

\bibitem[{\citenamefont{Popmintchev et~al.}(2009)\citenamefont{Popmintchev,
  Chen, Bahabad, Gerrity, Sidorenko, Cohen, Christov, Murnane, and
  Kapteyn}}]{KM:Phase_Matching}
\bibinfo{author}{\bibfnamefont{T.}~\bibnamefont{Popmintchev}},
  \bibinfo{author}{\bibfnamefont{M.-C.} \bibnamefont{Chen}},
  \bibinfo{author}{\bibfnamefont{A.}~\bibnamefont{Bahabad}},
  \bibinfo{author}{\bibfnamefont{M.}~\bibnamefont{Gerrity}},
  \bibinfo{author}{\bibfnamefont{P.}~\bibnamefont{Sidorenko}},
  \bibinfo{author}{\bibfnamefont{O.}~\bibnamefont{Cohen}},
  \bibinfo{author}{\bibfnamefont{I.~P.} \bibnamefont{Christov}},
  \bibinfo{author}{\bibfnamefont{M.~M.} \bibnamefont{Murnane}},
  \bibnamefont{and} \bibinfo{author}{\bibfnamefont{H.~C.}
  \bibnamefont{Kapteyn}}, \bibinfo{journal}{Proc. Natl. Acad. Sci. USA}
  \textbf{\bibinfo{volume}{106}}, \bibinfo{pages}{10516}
  (\bibinfo{year}{2009}).

\bibitem[{\citenamefont{Schmidt et~al.}(2010)\citenamefont{Schmidt, B\'{e}jot,
  Gigu\`{e}re, Shiner, Trallero-Herrero, \'{E}ric Bisson, Kasparian, Wolf,
  Villeneuve, Kieffer et~al.}}]{Bruno:Iddler_Compression}
\bibinfo{author}{\bibfnamefont{B.~E.} \bibnamefont{Schmidt}},
  \bibinfo{author}{\bibfnamefont{P.}~\bibnamefont{B\'{e}jot}},
  \bibinfo{author}{\bibfnamefont{M.}~\bibnamefont{Gigu\`{e}re}},
  \bibinfo{author}{\bibfnamefont{A.~D.} \bibnamefont{Shiner}},
  \bibinfo{author}{\bibfnamefont{C.}~\bibnamefont{Trallero-Herrero}},
  \bibinfo{author}{\bibnamefont{\'{E}ric Bisson}},
  \bibinfo{author}{\bibfnamefont{J.}~\bibnamefont{Kasparian}},
  \bibinfo{author}{\bibfnamefont{J.-P.} \bibnamefont{Wolf}},
  \bibinfo{author}{\bibfnamefont{D.~M.} \bibnamefont{Villeneuve}},
  \bibinfo{author}{\bibfnamefont{J.-C.} \bibnamefont{Kieffer}},
  \bibnamefont{et~al.}, \bibinfo{journal}{Applied Physics Letters}
  \textbf{\bibinfo{volume}{96}}, \bibinfo{pages}{121109}
  (\bibinfo{year}{2010}).

\bibitem[{\citenamefont{Yudin and Ivanov}(2001)}]{Yudin-Ivanov_Ionization}
\bibinfo{author}{\bibfnamefont{G.~L.} \bibnamefont{Yudin}} \bibnamefont{and}
  \bibinfo{author}{\bibfnamefont{M.~Y.} \bibnamefont{Ivanov}},
  \bibinfo{journal}{Phys. Rev. A} \textbf{\bibinfo{volume}{64}},
  \bibinfo{pages}{013409} (\bibinfo{year}{2001}).

\bibitem[{\citenamefont{Hankin et~al.}(2001)\citenamefont{Hankin, Villeneuve,
  Corkum, and Rayner}}]{intensity_calibration}
\bibinfo{author}{\bibfnamefont{S.}~\bibnamefont{Hankin}},
  \bibinfo{author}{\bibfnamefont{D.}~\bibnamefont{Villeneuve}},
  \bibinfo{author}{\bibfnamefont{P.}~\bibnamefont{Corkum}}, \bibnamefont{and}
  \bibinfo{author}{\bibfnamefont{D.}~\bibnamefont{Rayner}},
  \bibinfo{journal}{Phys. Rev. A} \textbf{\bibinfo{volume}{64}},
  \bibinfo{pages}{013405} (\bibinfo{year}{2001}).

\bibitem[{\citenamefont{Farrell et~al.}(2009)\citenamefont{Farrell, McFarland,
  Bucksbaum, and G\"{u}hr}}]{Gur:HHG_Calibration}
\bibinfo{author}{\bibfnamefont{J.~P.} \bibnamefont{Farrell}},
  \bibinfo{author}{\bibfnamefont{B.~K.} \bibnamefont{McFarland}},
  \bibinfo{author}{\bibfnamefont{P.~H.} \bibnamefont{Bucksbaum}},
  \bibnamefont{and} \bibinfo{author}{\bibfnamefont{M.}~\bibnamefont{G\"{u}hr}},
  \bibinfo{journal}{Opt. Express} \textbf{\bibinfo{volume}{17}},
  \bibinfo{pages}{15134} (\bibinfo{year}{2009}).

\bibitem[{\citenamefont{Ferrari et~al.}(2010)\citenamefont{Ferrari, Calegari,
  Lucchini, Vozzi, Stagira, Sansone, and Nisoli}}]{nisoli:high_energy_atto}
\bibinfo{author}{\bibfnamefont{F.}~\bibnamefont{Ferrari}},
  \bibinfo{author}{\bibfnamefont{F.}~\bibnamefont{Calegari}},
  \bibinfo{author}{\bibfnamefont{M.}~\bibnamefont{Lucchini}},
  \bibinfo{author}{\bibfnamefont{C.}~\bibnamefont{Vozzi}},
  \bibinfo{author}{\bibfnamefont{S.}~\bibnamefont{Stagira}},
  \bibinfo{author}{\bibfnamefont{G.}~\bibnamefont{Sansone}}, \bibnamefont{and}
  \bibinfo{author}{\bibfnamefont{M.}~\bibnamefont{Nisoli}},
  \bibinfo{journal}{Nat. Photon.} \textbf{\bibinfo{volume}{4}},
  \bibinfo{pages}{875} (\bibinfo{year}{2010}).

\bibitem[{\citenamefont{Shiner et~al.}(2011)\citenamefont{Shiner, Schmidt,
  Trallero-Herrero, Worner, Patchkovskii, Corkum, Kieffer, Legare, and
  Villeneuve}}]{Shiner:2e_Xe}
\bibinfo{author}{\bibfnamefont{A.~D.} \bibnamefont{Shiner}},
  \bibinfo{author}{\bibfnamefont{B.~E.} \bibnamefont{Schmidt}},
  \bibinfo{author}{\bibfnamefont{C.}~\bibnamefont{Trallero-Herrero}},
  \bibinfo{author}{\bibfnamefont{H.~J.} \bibnamefont{Worner}},
  \bibinfo{author}{\bibfnamefont{S.}~\bibnamefont{Patchkovskii}},
  \bibinfo{author}{\bibfnamefont{P.~B.} \bibnamefont{Corkum}},
  \bibinfo{author}{\bibfnamefont{J.-C.} \bibnamefont{Kieffer}},
  \bibinfo{author}{\bibfnamefont{F.}~\bibnamefont{Legare}}, \bibnamefont{and}
  \bibinfo{author}{\bibfnamefont{D.~M.} \bibnamefont{Villeneuve}},
  \bibinfo{journal}{Nat. Phys.} \textbf{\bibinfo{volume}{7}},
  \bibinfo{pages}{464} (\bibinfo{year}{2011}).

\bibitem[{\citenamefont{Jin et~al.}(submitted)\citenamefont{Jin, Le,
  Trallero-Herrero, and Lin}}]{Jin:Xe_Atto_HHG}
\bibinfo{author}{\bibfnamefont{C.}~\bibnamefont{Jin}},
  \bibinfo{author}{\bibfnamefont{A.-T.} \bibnamefont{Le}},
  \bibinfo{author}{\bibfnamefont{C.~A.} \bibnamefont{Trallero-Herrero}},
  \bibnamefont{and} \bibinfo{author}{\bibfnamefont{C.~D.} \bibnamefont{Lin}},
  \bibinfo{journal}{Phys. Rev. A}  (\bibinfo{year}{submitted}).

\bibitem[{\citenamefont{Jin et~al.}(2009)\citenamefont{Jin, Le, and
  Lin}}]{Jin:PRA2009}
\bibinfo{author}{\bibfnamefont{C.}~\bibnamefont{Jin}},
  \bibinfo{author}{\bibfnamefont{A.-T.} \bibnamefont{Le}}, \bibnamefont{and}
  \bibinfo{author}{\bibfnamefont{C.~D.} \bibnamefont{Lin}},
  \bibinfo{journal}{Phys. Rev. A} \textbf{\bibinfo{volume}{79}},
  \bibinfo{pages}{053413} (\bibinfo{year}{2009}).

\bibitem[{\citenamefont{Jin et~al.}(2011{\natexlab{a}})\citenamefont{Jin, Le,
  and Lin}}]{Jin:PRA2011}
\bibinfo{author}{\bibfnamefont{C.}~\bibnamefont{Jin}},
  \bibinfo{author}{\bibfnamefont{A.-T.} \bibnamefont{Le}}, \bibnamefont{and}
  \bibinfo{author}{\bibfnamefont{C.~D.} \bibnamefont{Lin}},
  \bibinfo{journal}{Phys. Rev. A} \textbf{\bibinfo{volume}{83}},
  \bibinfo{pages}{023411} (\bibinfo{year}{2011}{\natexlab{a}}).

\bibitem[{\citenamefont{Jin et~al.}(2011{\natexlab{b}})\citenamefont{Jin,
  W\"{o}rner, Tosa, Le, Bertrand, Lucchese, Corkum, Villeneuve, and
  Lin}}]{Jin:JPB2011}
\bibinfo{author}{\bibfnamefont{C.}~\bibnamefont{Jin}},
  \bibinfo{author}{\bibfnamefont{H.~J.} \bibnamefont{W\"{o}rner}},
  \bibinfo{author}{\bibfnamefont{V.}~\bibnamefont{Tosa}},
  \bibinfo{author}{\bibfnamefont{A.-T.} \bibnamefont{Le}},
  \bibinfo{author}{\bibfnamefont{J.~B.} \bibnamefont{Bertrand}},
  \bibinfo{author}{\bibfnamefont{R.~R.} \bibnamefont{Lucchese}},
  \bibinfo{author}{\bibfnamefont{P.~B.} \bibnamefont{Corkum}},
  \bibinfo{author}{\bibfnamefont{D.~M.} \bibnamefont{Villeneuve}},
  \bibnamefont{and} \bibinfo{author}{\bibfnamefont{C.~D.} \bibnamefont{Lin}},
  \bibinfo{journal}{J. Phys. B: At. Mol. Opt. Phys.}
  \textbf{\bibinfo{volume}{44}}, \bibinfo{pages}{095601}
  (\bibinfo{year}{2011}{\natexlab{b}}).

\bibitem[{\citenamefont{Lin et~al.}(2010)\citenamefont{Lin, Le, Chen,
  Morishita, and Lucchese}}]{CDLin:JPB-Review}
\bibinfo{author}{\bibfnamefont{C.~D.} \bibnamefont{Lin}},
  \bibinfo{author}{\bibfnamefont{A.-T.} \bibnamefont{Le}},
  \bibinfo{author}{\bibfnamefont{Z.}~\bibnamefont{Chen}},
  \bibinfo{author}{\bibfnamefont{T.}~\bibnamefont{Morishita}},
  \bibnamefont{and} \bibinfo{author}{\bibfnamefont{R.}~\bibnamefont{Lucchese}},
  \bibinfo{journal}{J. Phys. B: At. Mol. Opt. Phys.}
  \textbf{\bibinfo{volume}{43}}, \bibinfo{pages}{122001}
  (\bibinfo{year}{2010}).

\bibitem[{\citenamefont{Morishita et~al.}(2008)\citenamefont{Morishita, Le,
  Chen, and Lin}}]{Toru:PRL2008}
\bibinfo{author}{\bibfnamefont{T.}~\bibnamefont{Morishita}},
  \bibinfo{author}{\bibfnamefont{A.-T.} \bibnamefont{Le}},
  \bibinfo{author}{\bibfnamefont{Z.}~\bibnamefont{Chen}}, \bibnamefont{and}
  \bibinfo{author}{\bibfnamefont{C.~D.} \bibnamefont{Lin}},
  \bibinfo{journal}{Phys. Rev. Lett.} \textbf{\bibinfo{volume}{100}},
  \bibinfo{pages}{013903} (\bibinfo{year}{2008}).

\bibitem[{\citenamefont{Le et~al.}(2009)\citenamefont{Le, Lucchese, Tonzani,
  Morishita, and Lin}}]{ATLe:QRS2009}
\bibinfo{author}{\bibfnamefont{A.-T.} \bibnamefont{Le}},
  \bibinfo{author}{\bibfnamefont{R.~R.} \bibnamefont{Lucchese}},
  \bibinfo{author}{\bibfnamefont{S.}~\bibnamefont{Tonzani}},
  \bibinfo{author}{\bibfnamefont{T.}~\bibnamefont{Morishita}},
  \bibnamefont{and} \bibinfo{author}{\bibfnamefont{C.~D.} \bibnamefont{Lin}},
  \bibinfo{journal}{Phys. Rev. A} \textbf{\bibinfo{volume}{80}},
  \bibinfo{pages}{013401} (\bibinfo{year}{2009}).

\bibitem[{\citenamefont{Becker et~al.}(1989)\citenamefont{Becker, Szostak,
  Kerkhoff, Kupsch, Langer, Wehlitz, Yagishita, and Hayaishi}}]{Becker:PICS}
\bibinfo{author}{\bibfnamefont{U.}~\bibnamefont{Becker}},
  \bibinfo{author}{\bibfnamefont{D.}~\bibnamefont{Szostak}},
  \bibinfo{author}{\bibfnamefont{H.~G.} \bibnamefont{Kerkhoff}},
  \bibinfo{author}{\bibfnamefont{M.}~\bibnamefont{Kupsch}},
  \bibinfo{author}{\bibfnamefont{B.}~\bibnamefont{Langer}},
  \bibinfo{author}{\bibfnamefont{R.}~\bibnamefont{Wehlitz}},
  \bibinfo{author}{\bibfnamefont{A.}~\bibnamefont{Yagishita}},
  \bibnamefont{and} \bibinfo{author}{\bibfnamefont{T.}~\bibnamefont{Hayaishi}},
  \bibinfo{journal}{Phys. Rev. A} \textbf{\bibinfo{volume}{39}},
  \bibinfo{pages}{3902} (\bibinfo{year}{1989}).

\bibitem[{\citenamefont{Fahlman et~al.}(1984)\citenamefont{Fahlman, Krause,
  Carlson, and Svensson}}]{Fahlman:PICS}
\bibinfo{author}{\bibfnamefont{A.}~\bibnamefont{Fahlman}},
  \bibinfo{author}{\bibfnamefont{M.~O.} \bibnamefont{Krause}},
  \bibinfo{author}{\bibfnamefont{T.~A.} \bibnamefont{Carlson}},
  \bibnamefont{and} \bibinfo{author}{\bibfnamefont{A.}~\bibnamefont{Svensson}},
  \bibinfo{journal}{Phys. Rev. A} \textbf{\bibinfo{volume}{30}},
  \bibinfo{pages}{812} (\bibinfo{year}{1984}).

\bibitem[{\citenamefont{Kutzner et~al.}(1989)\citenamefont{Kutzner,
  Radojevi\'c, and Kelly}}]{Kutzner:PICS}
\bibinfo{author}{\bibfnamefont{M.}~\bibnamefont{Kutzner}},
  \bibinfo{author}{\bibfnamefont{V.}~\bibnamefont{Radojevi\'c}},
  \bibnamefont{and} \bibinfo{author}{\bibfnamefont{H.~P.} \bibnamefont{Kelly}},
  \bibinfo{journal}{Phys. Rev. A} \textbf{\bibinfo{volume}{40}},
  \bibinfo{pages}{5052} (\bibinfo{year}{1989}).

\bibitem[{\citenamefont{Zhou et~al.}(1996)\citenamefont{Zhou, Peatros, Murnane,
  and Kapteyn}}]{KM:HHG_Chirp}
\bibinfo{author}{\bibfnamefont{J.}~\bibnamefont{Zhou}},
  \bibinfo{author}{\bibfnamefont{J.}~\bibnamefont{Peatros}},
  \bibinfo{author}{\bibfnamefont{M.~M.} \bibnamefont{Murnane}},
  \bibnamefont{and} \bibinfo{author}{\bibfnamefont{H.~C.}
  \bibnamefont{Kapteyn}}, \bibinfo{journal}{Phys. Rev. Lett.}
  \textbf{\bibinfo{volume}{76}}, \bibinfo{pages}{752} (\bibinfo{year}{1996}).

\bibitem[{\citenamefont{Kim et~al.}(2004)\citenamefont{Kim, Kim, Hong, Lee,
  Kim, and Nam}}]{Nam:HHG_Chirp}
\bibinfo{author}{\bibfnamefont{H.~T.} \bibnamefont{Kim}},
  \bibinfo{author}{\bibfnamefont{I.~J.} \bibnamefont{Kim}},
  \bibinfo{author}{\bibfnamefont{K.-H.} \bibnamefont{Hong}},
  \bibinfo{author}{\bibfnamefont{D.~G.} \bibnamefont{Lee}},
  \bibinfo{author}{\bibfnamefont{J.-H.} \bibnamefont{Kim}}, \bibnamefont{and}
  \bibinfo{author}{\bibfnamefont{C.~H.} \bibnamefont{Nam}},
  \bibinfo{journal}{J. Phy. B: At. Mol. Opt. Phys.}
  \textbf{\bibinfo{volume}{37}}, \bibinfo{pages}{1141} (\bibinfo{year}{2004}).

\end{thebibliography}

\end{document}